\documentclass[reprint,aps,epsfig,superscriptaddress,twocolumn]{revtex4}
\usepackage{bm}
\usepackage{amsfonts}
\usepackage[dvips]{graphicx}
\usepackage{mathrsfs}
\usepackage[intlimits]{amsmath}
\usepackage{textcomp}
\usepackage[colorlinks, citecolor=red]{hyperref}
\usepackage{subfigure}
\usepackage{setspace}
\usepackage{tabularx}

\begin{document}

\title{Multiplexed ion-ion entanglement over $1.2$ kilometer fibers}

\author{Z.-B. Cui}
\altaffiliation{These authors contributed equally to this work}
\affiliation{Center for Quantum Information, Institute for Interdisciplinary Information Sciences, Tsinghua University, Beijing 100084, PR China}

\author{Z.-Q. Wang}
\altaffiliation{These authors contributed equally to this work}
\affiliation{Center for Quantum Information, Institute for Interdisciplinary Information Sciences, Tsinghua University, Beijing 100084, PR China}

\author{P.-Y. Liu}
\affiliation{Center for Quantum Information, Institute for Interdisciplinary Information Sciences, Tsinghua University, Beijing 100084, PR China}

\author{Y. Wang}
\affiliation{HYQ Co., Ltd., Beijing 100176, PR China}

\author{P.-C. Lai}
\affiliation{Center for Quantum Information, Institute for Interdisciplinary Information Sciences, Tsinghua University, Beijing 100084, PR China}

\author{J.-X. Shi}
\affiliation{Center for Quantum Information, Institute for Interdisciplinary Information Sciences, Tsinghua University, Beijing 100084, PR China}

\author{Y.-D. Sun}
\affiliation{Center for Quantum Information, Institute for Interdisciplinary Information Sciences, Tsinghua University, Beijing 100084, PR China}

\author{Z.-C. Tian}
\affiliation{Center for Quantum Information, Institute for Interdisciplinary Information Sciences, Tsinghua University, Beijing 100084, PR China}

\author{H.-S. Sun}
\affiliation{Center for Quantum Information, Institute for Interdisciplinary Information Sciences, Tsinghua University, Beijing 100084, PR China}

\author{Y.-B. Liang}
\affiliation{Center for Quantum Information, Institute for Interdisciplinary Information Sciences, Tsinghua University, Beijing 100084, PR China}

\author{B.-X. Qi}
\affiliation{Center for Quantum Information, Institute for Interdisciplinary Information Sciences, Tsinghua University, Beijing 100084, PR China}

\author{Y.-Y. Huang}
\affiliation{Center for Quantum Information, Institute for Interdisciplinary Information Sciences, Tsinghua University, Beijing 100084, PR China}

\author{Z.-C. Zhou}
\affiliation{Center for Quantum Information, Institute for Interdisciplinary Information Sciences, Tsinghua University, Beijing 100084, PR China}
\affiliation{Hefei National Laboratory, Hefei 230088, PR China}

\author{Y.-K. Wu}
\affiliation{Center for Quantum Information, Institute for Interdisciplinary Information Sciences, Tsinghua University, Beijing 100084, PR China}
\affiliation{Hefei National Laboratory, Hefei 230088, PR China}

\author{Y. Xu}
\affiliation{Center for Quantum Information, Institute for Interdisciplinary Information Sciences, Tsinghua University, Beijing 100084, PR China}
\affiliation{Hefei National Laboratory, Hefei 230088, PR China}

\author{Y.-F. Pu}
\email{puyf@tsinghua.edu.cn}
\affiliation{Center for Quantum Information, Institute for Interdisciplinary Information Sciences, Tsinghua University, Beijing 100084, PR China}
\affiliation{Hefei National Laboratory, Hefei 230088, PR China}

\author{L.-M. Duan}
\email{lmduan@tsinghua.edu.cn}
\affiliation{Center for Quantum Information, Institute for Interdisciplinary Information Sciences, Tsinghua University, Beijing 100084, PR China}
\affiliation{Hefei National Laboratory, Hefei 230088, PR China}

\begin{abstract}
Quantum networks and quantum repeaters represent the promising avenues for building large-scale quantum information systems, serving as foundational infrastructure for distributed quantum computing, long-distance quantum communication, and networked quantum sensing. A critical step in realizing a functional quantum network is the efficient and high-fidelity establishment of heralded entanglement between remote quantum nodes. Multiplexing offers a powerful strategy to accelerate remote entanglement distribution, particularly over long optical fibers. Here, we demonstrate the first multiplexing-enhanced heralded entanglement between two trapped-ion quantum network nodes. By multiplexing $10$ temporal photonic modes, we achieve a 4.59-fold speedup in ion-ion entanglement generation and attain an entanglement fidelity of $95.9\pm1.5\%$ over $1.2$ km of fiber. Employing a dual-type architecture, our system is readily scalable to multiple nodes, thereby establishing a key building block for future large-scale quantum networks.

\end{abstract}

\maketitle

Quantum networks and repeaters, which interconnect local quantum nodes via photonic channels, provide a leading pathway for scaling up quantum information systems. They serve as the backbone for numerous advanced applications, including long-distance quantum communication~\cite{BDCZ,DLCZ, duan arxiv,rmp_gisin,oxford_qkd,weifurner_qkd}, distributed quantum computing~\cite{distributed, duan arxiv, duan RMP, duan pra,oxford distributed, rempe}, and networked quantum sensing~\cite{ye_and_lukin,repeater_telescope, oxford_clock}. Among various physical platforms, trapped-ion systems have demonstrated exceptional performance in key metrics: they hold the record for the highest fidelity and rate in laboratory-scale heralded matter-matter entanglement~\cite{monroe_barium, monroe time-bin, oxford distributed}, the highest fidelities in quantum logic gates and state detection~\cite{oxford gate, oxford spam}, and among the longest coherence times~\cite{xuyulin, 1hour}. These outstanding properties establish trapped ions as a highly promising platform for building large-scale, multifunctional quantum networks.

A major bottleneck in realizing practical quantum networks is the limited efficiency and fidelity of the initial step---establishing heralded matter-matter entanglement between remote nodes---particularly over distances beyond the laboratory scale. Multiplexing presents a powerful strategy to enhance the entanglement distribution rate by effectively increasing the number of concurrent entangling attempts~\cite{2007prl, rmp_gisin, 2021nature, zhou, 1250,tittel_frequency, Oblak, oam,lan,225,multipurpose,enhancement,changwei,wavevector,lanyon multiplexing,haffner multiplexing, solid ion, cui, fiber array}. This approach is crucial for long-distance links, where the attempt rate is fundamentally limited by the round-trip communication time $L/c$ (with $L$ the fiber length and $c$ the speed of light in fiber). To date, multiplexing-enhanced matter-matter entanglement has been demonstrated only in solid-state systems~\cite{2021nature, zhou, solid ion}. Furthermore, the current record for matter-matter entanglement fidelity over fiber distances beyond the laboratory scale stands up to approximately $90\%$ for all physical platforms~\cite{230m, hanson bell state}.

In this work, we report the first experimental realization of multiplexing-enhanced matter-matter entanglement using a trapped-ion system. By sequentially utilizing $10$ temporal photonic modes from each ion network node, we achieve a $4.59$-fold enhancement in the ion-ion entangling efficiency compared to the single-mode case. The experiment is performed over a total fiber length of $1.2\,$km, surpassing the previous record of $520\,$m for trapped ions~\cite{230m}. Moreover, we achieve an ion-ion entanglement fidelity of $95.9\pm1.5\%$ over this distance. This result sets a new fidelity record for heralded matter-matter entanglement at extended fiber lengths across all physical platforms and is competitive with the best values achieved in laboratory-scale setups~\cite{oxford distributed, monroe time-bin}.

\begin{figure*}[]
    \centering
    \includegraphics[width=1.0\textwidth]{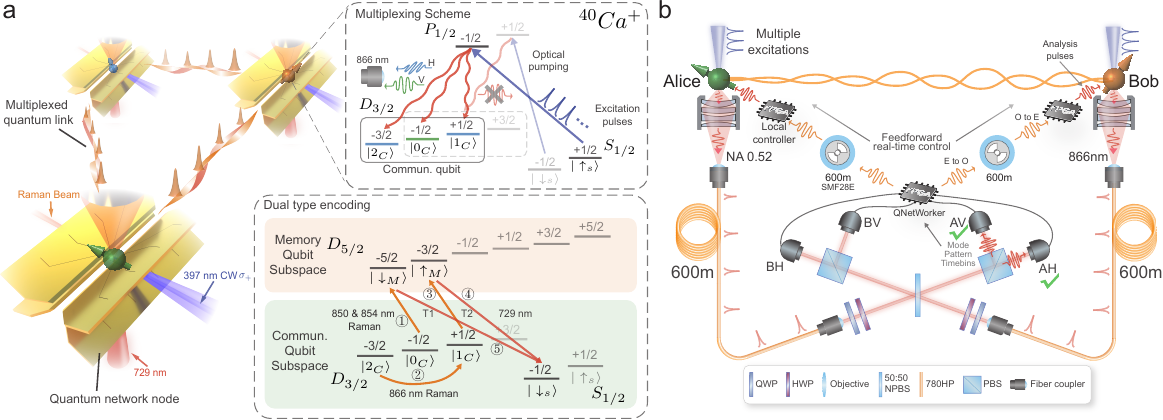}
    \caption{
    \textbf{Multiplexed trapped-ion quantum network.}
    \textbf{a}, The schematic of a multiplexed trapped-ion quantum network. Photonic interconnects are employed to link remote quantum network nodes, and multiplexing schemes are exploited to accelerate the remote entanglement generation. We use dual-type encoding to realize the trapped-ion quantum network node. Via the dual-type encoding, the lowest energy levels of a $^{40}$Ca$^+$ ion are splitted into two spectrally isolated subspaces, i.e., the communication qubit subspace and the memory qubit subspace, as shown in the lower inset ($|P_{1/2}\rangle$ is not shown here). The encoding conversions between the two subspaces and the qubit rotations are implemented by the Raman transitions and the $729\,$nm quadruple transition. The multiplexed ion-photon entanglement is generated by shining multiple $397\,$nm excitation pulses to the ion, as shown in the upper inset. We collect the $866\,$nm photons during the spontaneous emission of the excited state $|P_{1/2},m=-1/2\rangle$. All the cooling, pumping, ion-photon entanglement excitation, and state detection operations are performed in the communication subspace.
    \textbf{b}, The experimental setup of this work. The two quantum network nodes Alice and Bob each contain a $^{40}$Ca$^+$ inside, and a local controller is responsible for all the operations in each node. After the ion-photon entanglement generation, the emitted $866\,$nm photon from each node is transmitted through a $10\,$m or $600\,$m fiber to the measurement station in the center. The microprocessor `QNetWorker' decides whether a successful heralding is achieved based on the pattern of the photon detector clicks, and sends back the heralding signal to each node, via a fiber of the same length for the flying qubit. The local controller in each node then determines the next step based on the returned heralding signal.
    }
    \label{FIG1}
\end{figure*}

We build two trapped-ion quantum network nodes Alice and Bob based on $^{40}$Ca$^+$ ions, with one ion in each node, as illustrated in Fig.~\hyperref[FIG1]{1}. The ion is trapped in a segmented blade trap in each node and the two nodes are spatially separated by $2\,$m. We use the dual-type scheme for the realization of each quantum network node~\cite{yanghx, omg, huangyy,lai}. Via the dual-type encoding, the lowest energy levels of $^{40}$Ca$^+$ are split into two spectrally isolated groups, i.e., the communication qubit subspace which contains $|S_{1/2}\rangle$, $|D_{3/2}\rangle$, and $|P_{1/2}\rangle$, and the memory qubit subspace which contains $|D_{5/2}\rangle$, as shown in Fig.~\hyperref[FIG1]{1}a. The dissipative operations such as cooling, pumping, ion-photon entangling, and state detection in this experiment are performed in the communication qubit subspace which will not influence the memory qubit subspace due to the large spectral isolation. Although there is only one ion in each node in this experiment, we still use this dual-type framework to build the ionic quantum network, as it renders the ability to realize a multi-node quantum network with many qubits per node in the future~\cite{duan arxiv, lai, cui}.

As illustrated in Fig.~\hyperref[FIG1]{1}a, the $^{40}$Ca$^+$ ion in each node is first initialized in the state $|S_{1/2},m=+1/2\rangle$ via optical pumping. Then the ion is excited to $|P_{1/2},m=-1/2\rangle$ by a picosecond $397\,$nm laser pulse. The excited state $|P_{1/2},m=-1/2\rangle$ then spontaneously decays to three possible Zeeman states $|0_C\rangle\equiv|D_{3/2},m=-1/2\rangle$, $|1_C\rangle\equiv|D_{3/2},m=+1/2\rangle$, and $|2_C\rangle\equiv|D_{3/2},m=-3/2\rangle$ in the $|D_{3/2}\rangle$ level with a total branching ratio of $6\%$, and an $866\,$nm photon is emitted with different polarizations $\pi$, $\sigma^-$, and $\sigma^+$ in accordance with different decay channels. As the photon collection mode defined by the objective and the single mode fiber is perpendicular to the magnetic field, the circularly polarized photon will be projected to the $H$ polarization and the $\pi$-polarized photon will be projected to the $V$ polarization after being collected into the fiber. After three Raman pulses which combine $|1_C\rangle$ and $|2_C\rangle$ and convert the sub-levels from $|D_{3/2}\rangle$ to $|D_{5/2}\rangle$, the ion-photon state is prepared into a maximally entangled state
\begin{equation}
|\Phi\rangle_{\text{ion-photon}} =\frac{1}{\sqrt{2}}(|\uparrow_M\rangle|H\rangle+|\downarrow_M\rangle|V\rangle)
\end{equation}
where $|\uparrow_M\rangle\equiv|D_{5/2},m=-3/2\rangle$ and $|\downarrow_M\rangle\equiv|D_{5/2},m=-5/2\rangle$ are the two sub-levels in the memory qubit subspace. The detailed configuration of the photon excitation and the Raman pulses is described in Refs.~\cite{lai, cui} and the Supplementary Information.

We employ two-photon Bell-State Measurement (BSM) to herald the ion-ion entanglement, as shown in Fig.~\hyperref[FIG1]{1}b. Upon detecting a two-photon coincidence in the detectors AH and AV, the photonic state is projected into a Bell state $|\Psi^+\rangle=|H\rangle_\text{A}|V\rangle_\text{B}+|V\rangle_\text{A}|H\rangle_\text{B}$, and the heralded ion-ion entangled state is expressed as
\begin{equation}
|\Psi\rangle_{\text{ion-ion}} =\frac{1}{\sqrt{2}}(|\uparrow_M\rangle_\text{A}|\downarrow_M\rangle_\text{B}+e^{i\Delta\phi}|\downarrow_M\rangle_\text{A}|\uparrow_M\rangle_\text{B})
\label{IIE}
\end{equation}
where $\Delta\phi$ denotes the phase in the heralded ion-ion entanglement. A and B represent Alice and Bob, respectively.

\begin{figure}[]
    \centering
    \includegraphics[width=0.45\textwidth]{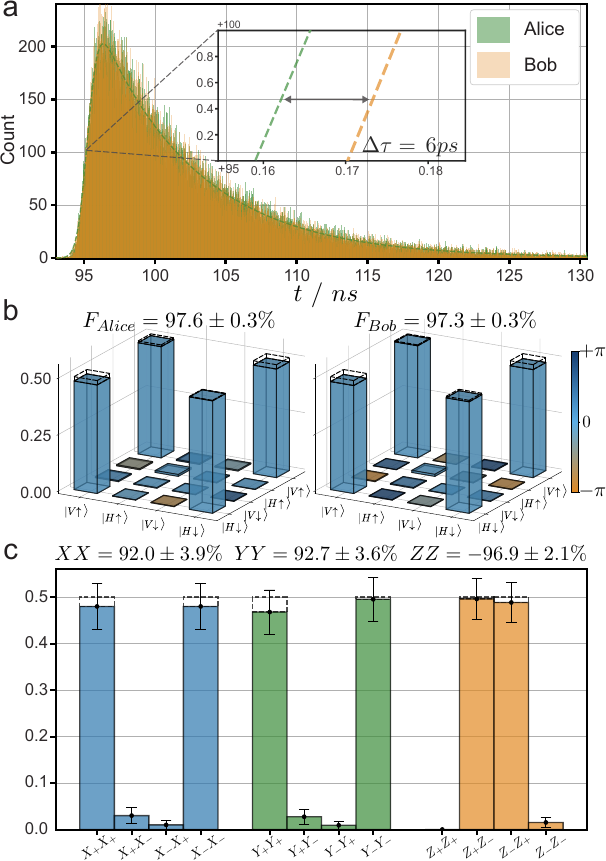}
    \caption{
    \textbf{Ion-ion entanglement over $20\,$m fibers.}
    \textbf{a}, The histogram for the arrival times of $866\,$nm photons from both nodes. This histogram is measured by the detectors in the middle measurement station. The two wavepackets for photons from Alice and Bob have a small shift of $6\,$ps after careful adjustments.
    \textbf{b}, The reconstructed density matrices of the ion-photon entangled states for both Alice and Bob.
    \textbf{c}, The measurement results of correlations XX, YY, and ZZ for the heralded ion-ion entangled state. Totally $340$ entanglement events are recorded for the measurements, with an ion-ion entangling rate of $0.039\,$s$^{-1}$.
    }
    \label{FIG2}
\end{figure}

We first demonstrate ion-ion entanglement over $20\,$m fibers. In this case, we only use the single mode excitation without multiplexing as the fiber length is short. In each remote entangling attempt, ion-photon entanglement is excited in each node and the emitted photon is sent through a $10\,$m fiber to the center station for measurement. To guarantee a high indistinguishability between the photons from two nodes, we carefully adjust the difference between the arrival times of photons from Alice and Bob, and achieve a small arrival time difference of $\Delta \tau=6\,$ps which is much shorter than the temporal length of the spontaneously emitted photon, as shown in Fig.~\hyperref[FIG2]{2}a. The fidelities of the ion-photon entanglement are measured to be $97.6\pm0.3\%$ and $97.3\pm0.3\%$ for Alice and Bob, respectively. The reconstructed density matrices via quantum state tomography are shown in Fig.~\hyperref[FIG2]{2}b. By projecting the photonic state into Bell state $|\Psi^+\rangle$ and carefully matching the magnetic fields in the two nodes, the phase of the heralded ion-ion entanglement $\Delta\phi$ is insensitive to the arrival time difference $\Delta t$ between the two heralding photons. All two-photon coincidence events within the full $45\,$ns detection window can be utilized without the need to select the coincidence events with a small $\Delta t$, which improves the heralding efficiency significantly. The detailed analysis of $\Delta\phi$ is described in the Supplementary Information. Finally, after the two-photon heralding, the fidelity of the heralded ion-ion entanglement with respect to the maximally entangled Bell state $|\Psi^+\rangle$ is obtained by measuring the correlations $\langle \sigma_i^A\sigma_i^B \rangle, i \in \{x,y,z\}$, which are simplified as XX, YY, and ZZ. We measure XX, YY, and ZZ via single-qubit rotations enabled by the $729\,$nm transitions and state detections. The measured correlations are shown in Fig.~\hyperref[FIG2]{2}c, yielding an ion-ion entanglement fidelity of $\mathcal{F}=\frac{1}{4}(1+\text{XX}+\text{YY}-\text{ZZ})=95.4\pm1.4\%$.

\begin{figure}[]
    \centering
    \includegraphics[width=0.48\textwidth]{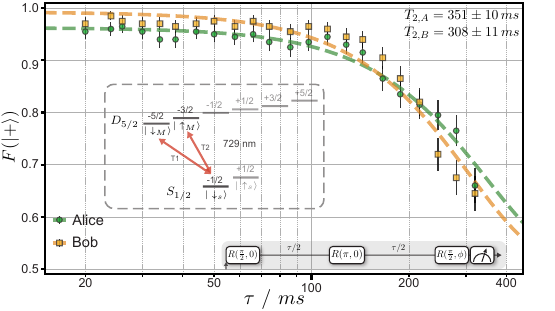}
    \caption{
    \textbf{Coherence time of each node.}
    We measure the coherence time of Alice and Bob by preparing the ion in a superposition state $|+\rangle=\frac{|\uparrow_M\rangle+|\downarrow_M\rangle}{\sqrt{2}}$ at first. Then a spin echo is performed in the middle of the storage duration $\tau$. After $\tau$, the stored state is characterized, and the fidelity with respect to the initial state is obtained. The fidelity decay of the stored state versus the storage time $\tau$ is demonstrated in the figure. The fitted coherence times for Alice and Bob are $351\pm10\,$ms and $308\pm11\,$ms, respectively.
    }
    \label{FIG3}
\end{figure}

We also measure the coherence time of each quantum network node. As the ionic qubit is encoded in $|\uparrow_M\rangle$ and $|\downarrow_M\rangle$, we characterize the memory coherence time by measuring the fidelity decay of the superposition state $|+\rangle=\frac{|\uparrow_M\rangle+|\downarrow_M\rangle}{\sqrt{2}}$ with increasing storage time. The time sequence for the coherence time measurement is illustrated in Fig.~\hyperref[FIG3]{3}. We first prepare $|+\rangle$ with a $\pi/2$-pulse of $729\,$nm transition T$_1$ followed by a $\pi$-pulse of transition T$_2$, starting from $|S_{1/2},m=-1/2\rangle$. After a waiting time of $\tau/2$, a spin echo operation is performed on the qubit via three $729\,$nm pulses. After another waiting time of $\tau/2$, the quantum state of the ion is characterized via $729\,$nm pulses and the following state detection. The coherence times for Alice and Bob are measured to be $T_{2,\,\text{A}}=351\pm10\,$ms and $T_{2,\,\text{B}}=308\pm11\,$ms, respectively.

\begin{figure*}[]
    \centering
    \includegraphics[width=\textwidth]{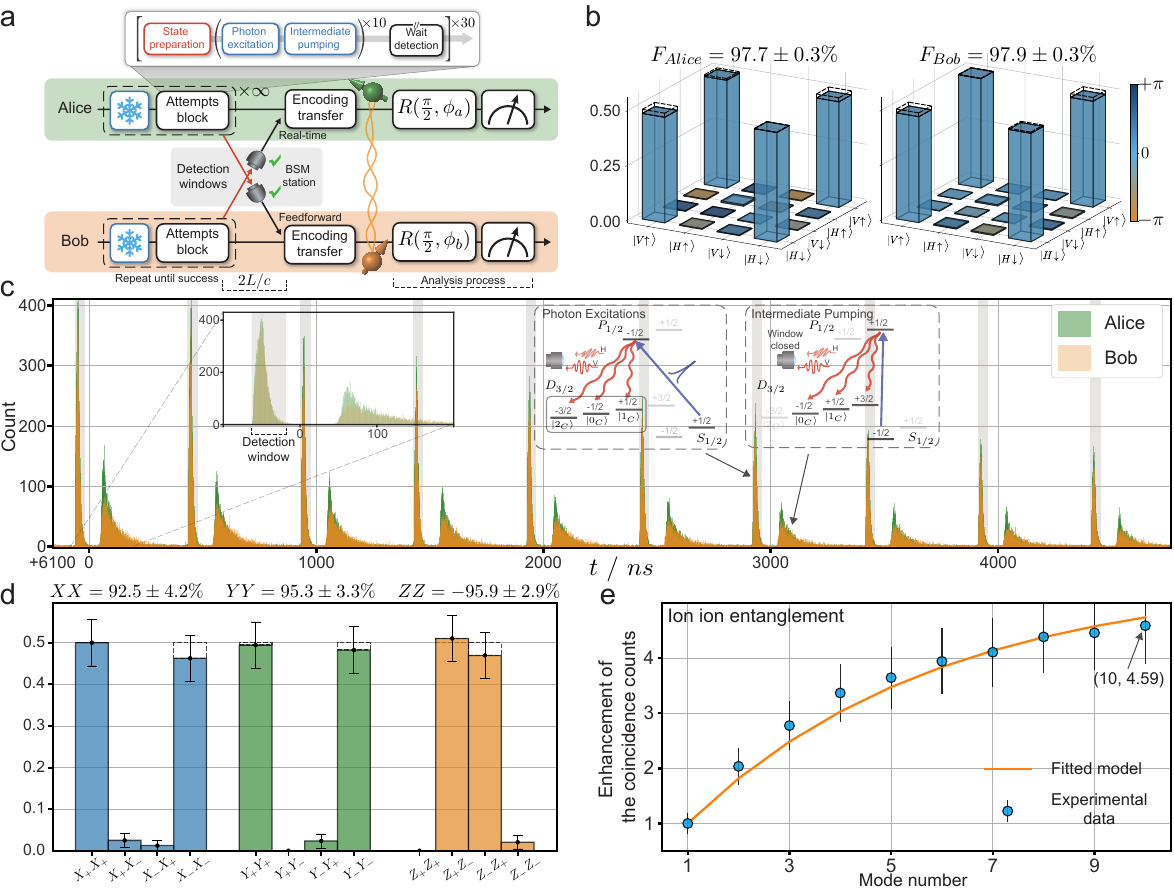}
    \caption{
    \textbf{Multiplexing-enhanced ion-ion entanglement over $1.2\,$km fibers.}
    \textbf{a}, The time sequence for the multiplexed ion-ion entanglement generation. The combination of an EIT cooling of $200\,\mu$s and a block of remote entangling attempts is continuously running until an ion-ion entanglement is successfully heralded. Totally $10\times30=300$ photon excitations are performed in each block of attempts.
    \textbf{b}, The reconstructed density matrices for the ion-photon entangled states from Alice and Bob. The photon state is measured in the middle measurement station.
    \textbf{c}, The histogram for the arrival time of the $866\,$nm photons recorded by the detectors in the middle station. The $10$ narrower peaks correspond to the photons emitted in the ion-photon entanglement excitations, and the detection window for the ion-ion entanglement heralding is $45\,$ns. The $10$ wider peaks are the $866\,$nm photons emitted during the intermediate pumping. Successive photon excitations have a time interval of $500\,$ns in between.
    \textbf{d}, The correlations XX, YY, and ZZ for measuring the fidelity of heralded ion-ion entanglement. Totally $263$ entanglement events are recorded for the measurements, with an ion-ion entangling rate of $0.011$s$^{-1}$.
    \textbf{e}, The enhancement factor of the multiplexed ion-ion entanglement over single-mode case. The efficiency enhancement versus the mode number used in each excitation round is shown here. The efficiency of the remote ion-ion entangling is enhanced by $4.59$ times with $10$ time-bin modes in each round.
    }
    \label{FIG4}
\end{figure*}

We employ the multiplexing scheme to enhance the ion-ion entangling efficiency in the $1.2\,$km case. During the spontaneous decay of the excited state $|P_{1/2},m=-1/2\rangle$, the ion decays to $|D_{3/2}\rangle$ level and emits an $866\,$nm photon with a low branching ratio of $6\%$, but decays back to the $|S_{1/2}\rangle$ level with a probability of $94\%$ (of which $2/3$ goes back to the initial state). This low branching ratio limits the efficiency of generating ion-photon entanglement with the $866\,$nm transition, which has a more favorable wavelength over the UV transition in quantum networking. To overcome the issue of low branching ratio, a multiplexing method of multiple excitation has been developed to effectively improve the branching ratio of $866\,$nm transition by reusing the remaining population in $|S_{1/2}\rangle$ level after prior excitations~\cite{cui}. Here we employ this multiplexing scheme to accelerate the ion-ion entanglement over a fiber length of $1.2\,$km. As illustrated in Fig.~\hyperref[FIG4]{4}a, the ion in each node is first cooled by a $200\,\mu$s EIT cooling, followed by a block of remote entangling attempts with $30$ rounds of photon excitations. Each excitation round contains $10$ successive photon excitations, and totally $30\times10=300$ entangling attempts are performed in each block. In each round of $10$ successive excitations, the ion is first initialized to $|S_{1/2},m=+1/2\rangle$ by a state preparation pulse with both $397\,$nm and $866\,$nm lasers, then $10$ consecutive `excitation-pumping' combinations are performed. In each intermediate optical pumping after a photon excitation, only the $397\,$nm laser is used. The $866\,$nm laser is not used to avoid destroying the state in $|D_{3/2}\rangle$ level if the prior photon excitation attempts were successful~\cite{cui}. The histogram of the photon arrival time during a round of $10$ successive excitations is measured (via the detectors in the center measurement station $600\,$m away from each node), as shown in Fig.~\hyperref[FIG4]{4}c. The time interval between two successive photon excitations is $500\,$ns, and each photon excitation is followed by an intermediate optical pumping of $350\,$ns.

The generated $866\,$nm photon from each node is transmitted through a $600\,$m fiber and is interfered in a $50$:$50$ beamsplitter at the center measurement station. The photon detection events in the single photon detectors are pre-processed by a microprocessor which we call `QNetWorker'. This microprocessor identifies the desired photon coincidence pattern corresponding to $|\Psi^+\rangle$ and then sends back the signal of whether the attempt is successful to the local control units of Alice and Bob. Here we use a strict heralding protocol that both the flying qubits (from local nodes to the center station) and the heralding signals (from the center station back to each node) are transmitted through the long fibers ($600\,$m in this case) to simulate the round-trip communication time in the real-world scenario. The transmission of the heralding signal back to local nodes is assisted by two E-O (electric-optical) converters, as shown in Fig.~\hyperref[FIG1]{1}b. With this strict heralding protocol, the time delay between a photon excitation attempt and the arrival of its heralding signal is approximately $\frac{2\times600\,\text{m}}{c}=6\,\mu$s in each local node, where $c$ is the light speed in fiber.

Here we still use a $45\,$ns detection window for ion-ion entanglement heralding, as shown in Fig.~\hyperref[FIG4]{4}c. If the two detectors corresponding to the desired photonic Bell state both record a click in the $45\,$ns detection window, we identify this as a successful heralding event, as in the $20\,$m case. This $45\,$ns detection window collects the vast majority of the incoming photons as the $1/e$ lifetime of the exponentially-decaying photon envelope is roughly $7\,$ns. In each round of $10$ successive excitations, the $10$ `excitation-pumping' operations cost about $5\,\mu$s in total (Fig.~\hyperref[FIG4]{4}c), thus the execution of an entire round of $10$ excitations is already finished when the heralding signal for the first mode returns to the local nodes, as the round-trip delay for the long fiber is $6\,\mu$s. After receiving the heralding signal, three successive Raman pulses are applied on the ion in each node to implement the encoding transfer, and the ion-ion entanglement as shown in Eq.~(\hyperref[IIE]{2}) is established. Finally, we perform local rotations and state detections on both ions to characterize the ion-ion entanglement fidelity with respect to the Bell state $|\Psi^+\rangle$, via the measurements of three correlations XX, YY, and ZZ. Thanks to the phase-insensitive heralding and feedforward ability, the same ion-ion entangled state is yielded at the end of the protocol no matter which mode is excited. The measurement results for the three correlation functions with all the $10$ time-bin modes combined are shown in Fig.~\hyperref[FIG4]{4}d, which guarantee an ion-ion entanglement fidelity of $\mathcal{F}=\frac{1}{4}(1+\text{XX}+\text{YY}-\text{ZZ})=95.9\pm1.5\%$.

Here we also investigate the enhancement factor of the ion-ion entangling efficiency through multiplexing. We find an enhancement of $4.59$ in ion-ion entangling efficiency is achieved with $10$ time-bin modes over single-mode case (only the first mode is excited in each round). The enhancement factor versus mode number is illustrated in Fig.~\hyperref[FIG4]{4}e, and it is shown that the multiplexing enhancement is nearly saturated with $10$ modes, as most of the population in $|S_{1/2}\rangle$ level has been squeezed to the $|D_{3/2}\rangle$ level after $10$ rounds of photon excitations and pumping. Here the ion-ion entanglement fidelity in the $1.2\,$km case is slightly higher than in the $20\,$m case. This is mainly due to the better cooling of the ion in the $1.2\,$km case. The error budgets for the ion-ion entanglement in both the $20\,$m and $1.2\,$km cases are listed in the Supplementary Information.

In conclusion, we have demonstrated the first multiplexing-enhanced entanglement between two trapped-ion quantum network nodes over $1.2\,$km of fiber. By employing temporal mode multiplexing, we achieved a $4.59$-fold enhancement in the entanglement distribution efficiency and a record-high fidelity of $95.9\%$ for long-distance matter-matter entanglement. Future work will focus on integrating additional multiplexing schemes---such as ion shuttling~\cite{lanyon multiplexing, haffner multiplexing, cui} and fiber arrays~\cite{fiber array}---with the multiple excitation method demonstrated here, targeting an enhancement factor exceeding $100$ using ion chains of tens of qubits. Further scaling of the fiber distance to tens of kilometers could be realized via quantum frequency conversion~\cite{weinfurter_33km, hanson long, lukin reflection, bao3nodes}. The current entanglement infidelity is primarily attributed to technical imperfections, including Raman detuning, phase errors, laser instability, insufficient cooling, and polarization mixing (see Supplementary Information). Addressing these issues could push the fidelity beyond $99\%$. Furthermore, the dual-type encoding scheme adopted here naturally supports scaling to multiple qubits per node, paving the way for multi-node networks ($>$$\,2$) capable of supporting advanced tasks such as entanglement purification, entanglement connection, and fully-functional quantum repeater nodes ~\cite{duan arxiv, lattice, hanson distillation, hanson 3 nodes}. This work thus establishes a high-fidelity, multiplexed, and long-distance ion-entanglement link as a critical building block for future large-scale quantum networks.

\noindent\textbf{Data availability:} The data that support the findings of this study are available from the
corresponding authors upon request.

\noindent\textbf{Acknowledgements:}
This work is supported by Innovation Program for Quantum Science and
Technology (No.2021ZD0301604, No.2021ZD0301102), the Tsinghua University
Initiative Scientific Research Program and the Ministry of
Education of China through its fund to the IIIS. Y.F.P. acknowledges
support from the Dushi Program
from Tsinghua University.

\noindent\textbf{Author Contributions:} Z.B.C., Z.Q.W., P.Y.L., Y.W., P.C.L., J.X.S., Y.D.S., Z.C.T., H.S.S., Y.B.L., B.X.Q., Y.Y.H., Z.C.Z., Y.K.W., Y.X. and Y.F.P. carried out the experiment. Y.F.P. and L.M.D. supervised the project. All the authors contributed to the discussion and the writing.

\noindent\textbf{Competing interests:} Y.K.W., Y.F.P., B.X.Q., Z.C.Z. and L.M.D. hold shares with the HYQ Co. The other authors declare no competing interests.

\noindent\textbf{Author Information:} The authors declare no competing financial interests.  Correspondence and requests for materials should be addressed to Y.F.P. (puyf@tsinghua.edu.cn) or L.M.D. (lmduan@tsinghua.edu.cn).

\onecolumngrid


\setcounter{equation}{0}
\setcounter{figure}{0}
\setcounter{table}{0}
\setcounter{page}{1}
\setcounter{section}{0}
\makeatletter

\renewcommand{\theHfigure}{S\arabic{figure}}
\renewcommand{\theHtable}{S\arabic{table}}
\renewcommand{\theHequation}{S\arabic{equation}}
\renewcommand{\theequation}{S\arabic{equation}}
\renewcommand{\thefigure}{S\arabic{figure}}
\renewcommand{\thetable}{S\arabic{table}}

\pagebreak
\begin{center}
\Large Supplementary Information for\\\textbf{``Multiplexed ion-ion entanglement over $1.2$ kilometer fibers''}
\end{center}

\subsection*{Section 1. Ion-photon entanglement generation and Raman transitions}

\begin{figure}[htbp]
    \centering
    \includegraphics[width=0.8\textwidth]{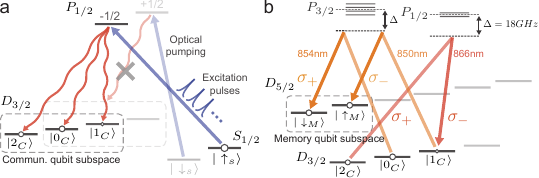}
    \caption{\textbf{Level scheme and transition diagram}. \textbf{a}, A $\sigma_+$ polarized $397\,$nm continuous-wave laser is employed for state preparation and intermediate pumping, and a $397\,$nm picosecond pulses transfers the ion from $|S_{1/2},m=+1/2\rangle$ to $|P_{1/2},m=-1/2\rangle$. We collect the $866\,$nm photons in the spontaneous decay to $|D_{3/2}\rangle$ level for the ion-photon entanglement. \textbf{b}, $850/854\,$nm Raman transition connects the $|D_{3/2}\rangle$ manifold in the communication qubit subspace and $|D_{5/2}\rangle$ manifold in the memory qubit subspace. $866/866\,\mathrm{nm}$ Raman transition is used for rotating $\frac{1}{2}|1_C\rangle + \frac{\sqrt{3}}{2} |2_C\rangle$ into $|1_C\rangle$ coherently.}
    \label{FIGS1}
\end{figure}

A single $794\,$nm pulse is picked by an electro-optical pulse picker from the pulse train produced by a mode-locked Ti:Sapphire laser with a repetition rate of $76\,$MHz. Then the single pulse passes through a second-harmonic generation (SHG) stage to generate the $397\,$nm pulse, which excites the ion from $|S_{1/2},m=+1/2\rangle$ to $|P_{1/2},m=-1/2\rangle$, as shown in Fig.~\hyperref[FIGS1]{S1}a.
After the ion spontaneously decays to the $|D_{3/2}\rangle$ manifold, the ion-photon state is expressed as
\begin{equation}
  |\Phi^+_{IP}\rangle = \frac{1}{\sqrt{3}} |0_C\rangle | \pi\rangle + \frac{1}{\sqrt{6}} |1_C\rangle | \sigma_-\rangle + \frac{1}{\sqrt{2}} |2_C\rangle | \sigma_+\rangle
\end{equation}

Since we collect the photons into fiber perpendicular to the magnetic field, the intensity of $\pi$ polarization is twice that of $\sigma$ polarization, and the $\pi$ and $\sigma$ polarizations are projected to $V$ and $H$ polarizations, respectively. Therefore, after the photon collection by a single-mode fiber, the ion-photon state is expressed as
\begin{equation}
  |\Phi^+_{IP}\rangle = \frac{1}{\sqrt{2}} |0_C\rangle |V\rangle + \frac{1}{\sqrt{2}} \left( \frac{1}{2}|1_C\rangle + \frac{\sqrt{3}}{2} |2_C\rangle \right) |H\rangle
\end{equation}

To obtain an ideal Bell-state, we apply three successive Raman pulses:
\begin{itemize}
  \item An $850/854\,\mathrm{nm}$ $\pi$-pulse, transfers $|0_C\rangle$ to $|\downarrow_M\rangle$.
  \item An $866/866\,\mathrm{nm}$ $\frac{2}{3}\ \pi$-pulse, combines $\frac{1}{2}|1_C\rangle + \frac{\sqrt{3}}{2} |2_C\rangle$ to $|1_C\rangle$.
  \item An $850/854\,\mathrm{nm}$ $\pi$-pulse, transfers $|1_C\rangle$ to $|\uparrow_M\rangle$.
\end{itemize}
After these Raman transitions, we end up with a maximally entangled Bell-state:
\begin{equation}
  |\Phi^+_{IP}\rangle = \frac{1}{\sqrt{2}} |\downarrow_M\rangle |V\rangle + \frac{1}{\sqrt{2}} |\uparrow_M\rangle |H\rangle
\end{equation}
\subsection*{Section 2. Phase analysis and experimental methodology}

\subsubsection*{1. Analysis of the phase in Bell state} \label{apbs}
To generate long-distance entanglement between two ions in separate traps connected by a $1.2\,$km optical channel,
we use $397\,$nm picosecond laser pulses to excite the ions from the state $|S_{1/2}, m=+1/2\rangle$ to the state $|P_{1/2}, m=-1/2\rangle$ within a few picoseconds, much shorter than the excited-state lifetime.
To reduce loss along the long optical path, we collect the $866\,$nm photon emitted from the decay to the $|D_{3/2}\rangle$ manifold using a high-numerical-aperture objective.
Because the branching ratio to the $|D_{3/2}\rangle$ manifold is relatively small, we repeat the excitation sequence multiple times to increase the entanglement generation rate.
Each decay channel from the state $|P_{1/2}, m=-1/2\rangle$ to the different $|D_{3/2}\rangle$ sublevels yields different photon polarization.
This naturally leads to entanglement between ion's internal state and the polarization of the emitted photon.
By interfering the photons emitted from Alice and Bob and performing a joint measurement, we project the two distant ions into an entangled state across the $1.2\,$km channel.

\begin{figure*}[htbp]
  \centering
  \includegraphics[width=1.0\textwidth]{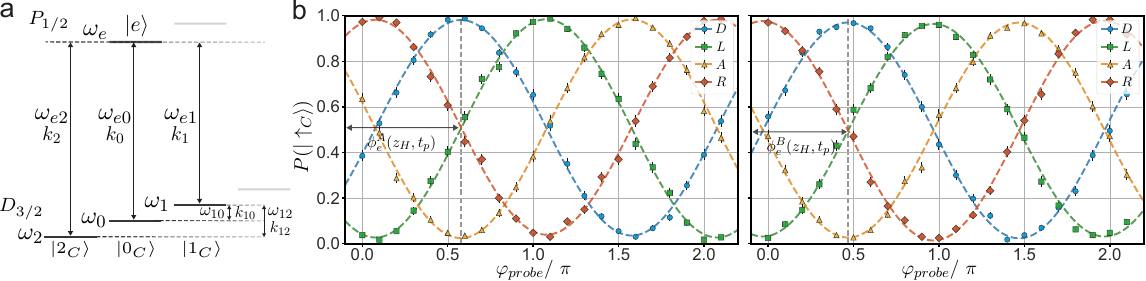}
  \caption{\textbf{Energy levels and phase calibration}. \textbf{a}, Energy levels and notations used in the analysis. \textbf{b}, The parity oscillation of the ion-photon entangled state encoded in the $|\downarrow_M\rangle$ and $|\uparrow_M\rangle$ is measured by scanning the phase of a $729\,$nm $\pi/2$ pulse applied on the qubit transition.}
  \label{FIGS2}
\end{figure*}

We begin our analysis by considering the ion-photon entangled state produced via spontaneous emission.
Spontaneous emission from the state $|P_{1/2}, m=-1/2\rangle$ arises from the dipole coupling with the vacuum electromagnetic field, and the resulting quantum state can be expressed as
\begin{equation}\label{IPE}
|\Phi^+_{IP}\rangle = \left(\sqrt{1/2}\,|0_C\rangle\,\tilde{a}^\dagger_{0V} +\sqrt{1/8}\,|1_C\rangle\,\tilde{a}^\dagger_{1H} +\sqrt{3/8}\,|2_C\rangle\,\tilde{a}^\dagger_{2H} \right) |0\rangle\,.
\end{equation}
One should notice that when collecting single photons perpendicular to the magnetic-field axis, the $\sigma_+$ and $\sigma_-$ transitions map to horizontal polarization, while the $\pi$ transition maps to vertical polarization.
Because the collection efficiency for $\sigma$ polarized light is half that for $\pi$ polarized light, the resulting ion-photon entangled state is a maximally entangled Bell state~\cite{lai}.
We filter out the spontaneous emission to the $|S_{1/2}\rangle$ manifold. Since the heralding detection of an $866\,$nm photon unambiguously projects the ion into the $|D_{3/2}\rangle$ manifold, no population is in $|S_{1/2}\rangle$ manifold in this case.
We define the photon creation operator as a normalized field operator with the spectral distribution of the emitted photon:
\begin{equation}\label{PCO}
\tilde{a}^\dagger_{i\varPi_i}(z,t)=\sqrt{\frac{\Gamma}{2\pi}}\int_{0}^\infty\frac{1}{\frac{\Gamma}{2}-i(\omega-\omega_{ei})}e^{i\omega\left(\frac{(z+z_0)}{c}-(t+t_0)\right)}a^\dagger_{\omega\varPi_i} \ d\omega\, ,
\end{equation}
where we define $\varPi_i$ as the polarization corresponding to the decay channel into the state $|i_C\rangle$, $\Gamma$ as the decay rate of the state $|P_{1/2}, m=-1/2\rangle$, $a^\dagger_{\omega\varPi_i}$ as the photon creation operator with frequency $\omega$ and polarization $\varPi_i$,
and $\omega_{ei}$ as the energy difference between the states $|P_{1/2}, m=-1/2\rangle$ and $|i_C\rangle$, as shown in Fig.~\hyperref[FIGS2]{S2}a.
The zero point of time ($t=0$) is set to the moment when the photon arrives at the beamsplitter (BS). The spatial zero point ($z=0$) is set to the position of BS. The ion is placed at $-z_0$, and the excitation pulse reaches it at time $-t_0$.
For the $V$ polarization, another $-1$ needs to be multiplied as the corresponding Clebsh-Gordan coefficients is a negative number.
The time-domain distribution of the spontaneously emitted photon is obtained by taking the Fourier transform of its spectral distribution, yielding an exponentially decaying envelope.

\begin{figure}[]
    \centering
    \includegraphics[width=0.65\textwidth]{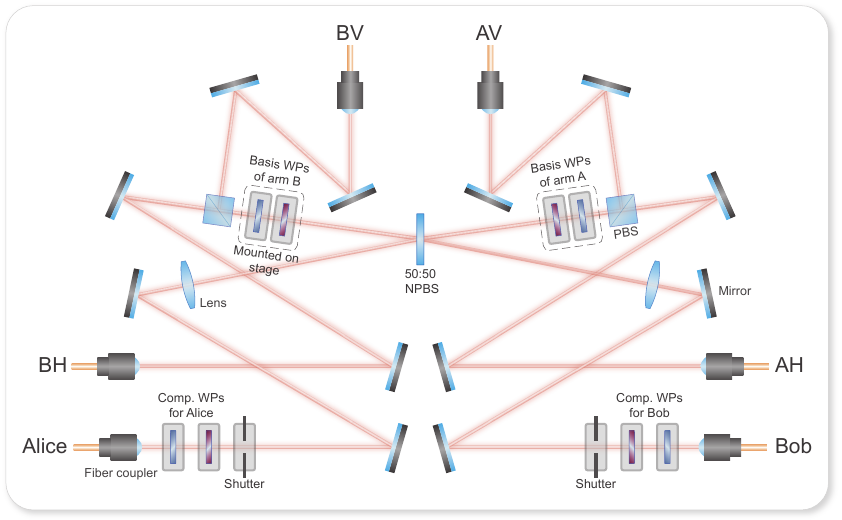}
    \caption{\textbf{The Bell-state measurement apparatus}.}
    \label{FIGS3}
\end{figure}

We now analyze the ion-ion entanglement.
The setup for Bell-state measurement (BSM) is shown in Fig.~\hyperref[FIGS3]{S3}.
Two polarization-encoded photons, emitted from Alice and Bob, arrive simultaneously at the BS and interfere with each other.
When the photons are highly indistinguishable, identical polarization components entering the BS from different ports always exit from the same output port, which is the well-known Hong-Ou-Mandel (HOM) effect.
Consequently, a coincidence detection of photons with orthogonal polarizations projects the two remote ions into an entangled state.
The joint ion-photon states of Alice and Bob can be decomposed into a combination of different Bell states:
\begin{equation}\label{DDW}
|\Psi\rangle_{AB} = {|\Phi^+_{IP}\rangle}_A \otimes {|\Phi^+_{IP}\rangle}_B = 1/2 \left( |\Phi^+_{II}\rangle \otimes |\Phi^+_{PP}\rangle + |\Phi^-_{II}\rangle \otimes |\Phi^-_{PP}\rangle + |\Psi^+_{II}\rangle \otimes |\Psi^+_{PP}\rangle + |\Psi^-_{II}\rangle \otimes |\Psi^-_{PP}\rangle\right)\,,
\end{equation}
where $|\Phi^\pm_{II}\rangle$ and $|\Psi^\pm_{II}\rangle$ denote the four Bell states of the two ions, and $|\Phi^\pm_{PP}\rangle$ and $|\Psi^\pm_{PP}\rangle$ represent the corresponding Bell states of the two photons.
Due to the HOM interference, photon pairs in the first two Bell states do not produce coincidence events, and thus only the latter two Bell states can be detected.
Consequently, the effective entanglement generation rate is given by $\eta = \eta_A \eta_B/2$, where $\eta_A$ and $\eta_B$ correspond to the efficiencies of the end-to-end links for Alice and Bob, respectively.
In the interference setup, four single-photon detectors (SPDs) are used, as illustrated in Fig.~\hyperref[FIGS3]{S3}.
They are labeled as AH, AV, BH, and BV, respectively, and the valid coincidence detection patterns are (AH, AV), (BH, BV), (AH, BV), (BH, AV)~\cite{iie_oxford}.

Ideally, the ion-ion entangled state possesses a fixed phase offset. However, drifts in the experimental system cause this offset to vary over time.
Such variations can influence the subsequent computational tasks if left uncorrected.
Thus, it is crucial to design a reliable method for calibrating and compensating this slow phase drift in the entangled state.
Here we first analyze the origin of this phase.
In Eq.~\eqref{IPE}, we obtain the ion-photon entangled state in a single network node.
After passing through the BS, the photon wave packet is split into two, acquiring fixed but distinct phase shifts in components corresponding to different input directions and polarizations.
Here, we consider a model of an asymmetric BS, whose transformation matrix can be expressed as
\begin{equation}\label{BS}
    U_{BS, \varPi}=
    \begin{bmatrix}
        t_\varPi & r_\varPi \\
        -{r_\varPi}^* & t_\varPi \\
    \end{bmatrix}\,.\\
\end{equation}
Here, $t_\varPi = \sqrt{1-R}$ and $r_\varPi = \sqrt{R}e^{i\phi^\varPi_{BS}}$ correspond to the complex transmission and reflection amplitudes, and $\varPi$ represents the polarization degree of freedom, with different phase shifts in different polarizations according to Fresnel's laws.

Taking the heralding event (AH, AV) with detection positions $(z_{H}, z_{V})$ and time $(\tau, \tau+\Delta\tau)$ as an example, through a straightforward derivation and combining with Eq.~\eqref{IPE} and Eq.~\eqref{PCO}, the product entangled state in Eq.~\eqref{DDW} collapses into the desired ion-ion entangled state, which can be expressed as
\begin{equation}\label{IIEP}
|\Psi^+_{II}\rangle = \alpha\,{|0_C\rangle}_A\otimes\left({\sqrt{1/4}|1_C\rangle}_B + \sqrt{3/4}\,e^{i\phi_{M}^B}{|2_C\rangle}_B\right)  + \beta\,e^{i\phi_E}\left({\sqrt{1/4}|1_C\rangle}_A + \sqrt{3/4}\,e^{i\phi_{M}^A}{|2_C\rangle}_A\right)\otimes{|0_C\rangle}_B\,.
\end{equation}
Here $\alpha$ and $\beta$ represent the amplitudes.

We define $\phi_E$ as the intermodular phase of the entangled state metioned above
\begin{equation}\label{phie}
\phi_E(t) = -\Delta\omega_{10}(t-\tau) - \Delta\omega_{e0} \Delta\tau + (\Delta k_1 z_H - \Delta k_0 z_V) + (k^A_{10} z^A_0 - k^B_{10} z^B_0) + (\phi^H_{BS} - \phi^V_{BS})\,.
\end{equation}
In the above equation, we define $\Delta \omega_{ij} = \omega^A_{ij}-\omega^B_{ij}$,  $\Delta k_i = (\omega^A_{ei}-\omega^B_{ei})/c$, $k^\alpha_{ij} = (\omega^\alpha_i-\omega^\alpha_j)/c$ and $z_\varPi$ refers to the position of the corresponding SPD.
The corresponding level structure is illustrated in  Fig.~\hyperref[FIGS2]{S2}a. Here, the indices $i$,$j$ denote the energy levels, $\alpha$ represents different ions, and $\varPi$ represents the polarization.
In the above expression, $\Delta \tau$ is completely random in each heralding, varying on the timescale of the detection window.
Consequently, the term involving $\Delta \tau$ can affect the fidelity.
In the experiment, by aligning the magnetic fields of the two nodes, we ensure that $\Delta\omega_{e0} \sim 0.2\,$kHz.
With a detection time window of approximately $50\,$ns, the error arising from this term is negligible.
The first term in the above equation also depends on time.
This term indicates that the Bell state phase evolves slowly over time due to the imperfect alignment of the magnetic fields.
It also constitutes the main source of dephasing for the long-distance entangled state, since fluctuations in the stray magnetic fields at each node affect the energy difference between the states ${|0_C\rangle}_\alpha$ and ${|1_C\rangle}_\alpha$ locally.
As shown in the main text, the individual coherence times of Alice and Bob are measured, as depicted in the main text.
The third and fourth terms arise from phase accumulation during photon propagation through the fiber, and fluctuations in the optical path can also lead to decoherence.
For the third term, with $2\pi/\Delta k_i \sim 10^{5}\,$m and the distance from the BS to the detectors is less than $2\,$m. This contribution can be safely neglected.
The fourth term is non-negligible, with $2\pi/\Delta k^\alpha_{10} \sim 43\,$m and $z_0^{\alpha} = 600\,$m, but phase offset is fixed and can be experimentally compensated.
The fidelity is affected by the optical path fluctuation $|\delta z_0^A| + |\delta z_0^B|$.
Using the thermal expansion coefficient of fused silica, we estimate that a $20\,$K temperature variation over a $1.2\,$km fiber results in a peak length change on the order of $10^{-2}\,$m.
This level of length variation induces negligible decoherence for our protocol.
In practice, the dominant instability arises from polarization drift caused by stress-induced birefringence in the fiber, which must therefore be stabilized.
The phase of the excitation laser does not appear explicitly in Eq.~\eqref{IIEP}.
In fact, for a two-photon scheme, the phase of the excitation pulses only contributes as a global phase.
This implies that the excitation pulses at the two nodes do not need to be phase-locked, which constitutes a significant engineering advantage.

We define $\phi_M$ as the phase between state ${|2_C\rangle}_\alpha$ and ${|1_C\rangle}_\alpha$. It can be written as
\begin{equation}\label{phim}
  \phi^\alpha_M(t) = \omega^\alpha_{12}(t-\tau) + k^\alpha_{12}(z_H+z^\alpha_0)\,.
\end{equation}
After the photon enters the fiber, the polarizations corresponding to the two states are both mapped to horizontal polarization.
We track the accumulated phase and then use an $866\,$nm Raman transition to coherently merge the two states.
It is worth noting that only the arrival time of the $H$ polarized photon determines the phase used to merge the two levels.
As seen from the phase $\phi_E$ of the Bell state above, as long as the magnetic fields of the two nodes are calibrated to be sufficiently equal to each other, the entangled state remains stationary or evolves only very slowly in time.
Consequently, subsequent operations exhibit robustness against uncertainties in their initiation times, ensuring that the entangled state can be directly utilized for further computational tasks once established.
Taken together, this means that in the experiment, only the merging operation needs to be referenced to the time of the $H$ polarized photon.

\subsubsection*{2. Phase calibration}
We can obtain the phase of ion-ion entanglement by calibrating the phases in the two ion-photon entanglement. Based on Eq.~\eqref{IPE},~\eqref{PCO}, and~\eqref{BS}, we can derive the phase of the ion-photon entangled states for Alice and Bob, respectively.
Assuming a perfect $50$:$50$ BS, the resulting ion-photon entangled state can be written as
\begin{equation}\label{IPEP}
{|\Phi^+_{IP}\rangle}_\alpha = 1/\sqrt{2}\left[{|0_C\rangle}_\alpha|V\rangle + e^{i\phi^\alpha_{e}} \left({\sqrt{1/4}|1_C\rangle}_\alpha + \sqrt{3/4}\,e^{i\phi_{M}^\alpha}{|2_C\rangle}_\alpha\right)|H\rangle\right]\,.
\end{equation}
The phase $\phi_{M}^\alpha$ between state ${|1_C\rangle}_\alpha$ and ${|2_C\rangle}_\alpha$ is identical to that in the ion-ion entangled state.
Therefore, in the experiment, this phase can be directly calibrated by the ion-photon entangled states.
For the phase $\phi^\alpha_{e}$ of the ion-photon entangled state, it can be expressed as
\begin{equation}\label{phieip}
\begin{aligned}
\phi^A_e(z, t) &= -\omega^A_{10}(t-\tau) + k^A_{10} (z + z^A_0) + \pi\,,\\
\phi^B_e(z, t) &= -\omega^B_{10}(t-\tau) + k^B_{10} (z + z^B_0) + \pi + (\phi^V_{BS} - \phi^H_{BS})\,.
\end{aligned}
\end{equation}
As depicted in Fig.~\hyperref[FIGS3]{S3}, in each arm of the BSM apparatus, we place a waveplate assembly (including a QWP and an HWP) mounted on a motorized translation stage, allowing it to be inserted or removed from the optical path.
After calibration using classical light, the combination of the waveplate assembly and the output ports of the PBS can be treated as a single photon projection measurement operator.
Fig.~\hyperref[FIGS2]{S2}b presents the results of phase scans of the $729\,$nm pulse performed by Alice and Bob in the four off-diagonal measurement bases.
A joint fit to the phases of these four curves allow us to extract $\phi^A_e(z_H, t_p)$ and $\phi^B_e(z_H, t_p)$. In this context, $t_p$ represents the moment at which the analysis process occurs.

From Eq.~\eqref{phie} and~\eqref{phieip}, one can readily observe the relationship between them:
\begin{equation}\label{relat}
  \phi_E(t)-\left[\phi^A_e(z_H, t)-\phi^B_e(z_H, t)\right] = -\Delta \omega_{e0}\Delta \tau + \Delta k_0(z_H-z_V) \approx 2\pi\times 10^{-5}\,.
\end{equation}
This result indicates that the intermodular phase in the ion-ion entangled state can be directly determined via the ion-photon entanglement.

\subsubsection*{3. State merging and encoding transfer}
As long-distance entanglement serves as a crucial computational resource for distributed quantum computing, it must be stored for subsequent computational tasks.
In our experiment, the states $|1_C\rangle$ and $|2_C\rangle$ are first merged using $866\,$nm Raman transitions to $|2_C\rangle$, and then states $|0_C\rangle$ and $|1_C\rangle$ are transferred to the states $|\downarrow_M\rangle$ and $|\uparrow_M\rangle$ in the memory qubit  subspace for storage, via $850/854\,$nm Raman transitions.
The details of this procedure are described in the prior sections and in~\cite{lai}.
According to Eq.~(\hyperref[phie]{S9}), the phase of the entangled state in the memory qubit subspace incorporates the phase $\phi_E(t_m)$ accumulated in the communication qubit subspace, as well as the phase progressively accumulated in memory qubit subspace due to the small magnetic field difference between the two quantum network nodes.
The entangled state after transfer to the memory qubit subspace can be expressed as

\begin{equation}\label{iie}
  {|\Psi\rangle}_{AB} = 1/\sqrt{2}\left({|\uparrow_M\downarrow_M\rangle}_{AB} + e^{i\Delta\phi} {|\downarrow_M\uparrow_M\rangle}_{AB}\right)\,.
\end{equation}
And the $\Delta\phi$ is given by
\begin{equation}\label{iiep}
  \Delta\phi = \Delta\omega_{\uparrow\downarrow}(t-t_m) + \Delta\omega_{10}t_m - \bar{k}_{10}\Delta z_0 - \phi_{misc}\,,
\end{equation}
where $\bar{k}_{10}$ is defined as $(\omega^A_{10}+\omega^B_{10})/2c$, $t_m$ denotes the timet at which the encoding transfer occurs, and \(\Delta z_0\) represents the fiber length difference between Alice's and Bob's optical channels to the BSM setup.
$\phi_{\mathrm{misc}}$ accounts for constant phase offsets arising from the Raman operations (e.g., light shifts) and the phase acquired at the BS.
We set the time origin to the detection of the $H$ polarized photon and neglect all terms that have been shown above to be negligible.
$\Delta\omega_{\uparrow\downarrow} = \omega^A_{\uparrow\downarrow} - \omega^B_{\uparrow\downarrow}$ and $\Delta\omega_{10} = \omega^A_{10} - \omega^B_{10}$ denote the energy differences between Alice's and Bob's qubits in the respective encoding bases.
\subsection*{Section 3. Data process and analysis}

\subsubsection*{1. Method for measuring entangled state}
The density matrix corresponding to an ideal Bell state can be expressed as
\begin{equation}\label{iiie}
|\Psi^+\rangle\langle\Psi^+| = 1/4\left(I\otimes I + X\otimes X + Y\otimes Y - Z\otimes Z\right)\,.
\end{equation}
The fidelity of the experimentally prepared two-ion state $\rho$ with respect to the ideal Bell state is defined as
\begin{equation}\label{f}
F\left(\rho, |\Psi^+\rangle\right) = \langle \Psi^+ | \rho | \Psi^+ \rangle = 1/4\left(1+\langle XX\rangle+\langle YY\rangle-\langle ZZ\rangle\right)\,.
\end{equation}

\subsubsection*{2. Infidelity analysis and error budget}
In this section, we analyze the sources of errors in our system.
The phase of the ion-ion entangled state has been thoroughly analyzed in the above sections,
where several error sources that can be neglected in our experiment are also discussed and quantitatively estimated.
Overall, the dominant errors mostly originate from technical imperfections such as imperfect laser configurations, stress-induced polarization drifts, and ion heating.
Below, we analyze some errors in the ion-ion entanglement generation.

The errors arising from the encoding transfer mainly originate from unwanted excitation and phase jitter in the $866\,$nm Raman merging operation.
These processes collectively lead to residual population remaining in the communication qubit subspace, which reduces the visibility of the $ZZ$ measurement and affects the visibility of the off-diagonal correlations.
The unwanted excitation in Raman induced by a single $\pi$ pulse can be estimated as $P_{sc} = 2\pi \Gamma/\Delta$.
From this, we estimate the infidelity induced by the Raman scattering is $\epsilon_{sc} = 0.98\%$.
The time jitter of the control pulses is measured to be approximately $3.3\,$ns.
We model the phase jitter of the $866\,$nm merging operation as following a Gaussian distribution with standard deviation $\sigma_\phi$, and the mean absolute deviation within one standard deviation is approximately $0.54\sigma_\phi$.
The residual population remaining in state $|2_C\rangle$ after the encoding transfer can be calculated as $(0.54\sigma_\phi)^2/2 = 0.55\%$, and the error induced by it is $\epsilon_{merge} = 0.41\%$.

In the experiment, we find that during the analysis of the entangled state using the $729\,$nm laser,
the $397\,$nm picosecond pulses and optical pumping operations induce heating of the ions, which significantly degrades the fidelity of single-qubit operations and consequently reduces the measured fidelity.
The relation between the visibility of the off-diagonal elements and the phonon number can be expressed as $V = 1 - \pi^2\eta^4(\bar{n}^2+\bar{n})\,$.
Combining the sequence parameters discussed in the main text, the average phonon numbers can be estimated for the $20\,$m and $1.2\,$km cases.
From these, the infidelity arising from motional heating in the two cases can be evaluated as $\epsilon^{20m}_{heat} = 0.83\%$ and $\epsilon^{1.2km}_{heat} = 0.31\%$.
It is worth noting that this contribution does not lead to the infidelity of the entangled state itself,
but induces errors during the quantum state analysis process.
In the $20\,$m case, we perform one cooling operation after every $300$ state preparation and photon excitation attempts. For the $1.2\,$km case, we employ ten temporal modes and repeat thirty rounds after a cooling operation, which is also one cooling operation every $300$ excitations.
However, two factors contribute to the reduced heating in the $1.2\,$km case.
First, as shown in the figure of the main text, the excitation probability of the ion decreases for the later modes in each round of $10$ excitations. Once the ion decays to $|D_{3/2}\rangle$ state in any of the $10$ modes (either due to photon excitation or intermediate pumping), the following `excitation-pumping' iterations in this round will no longer influence the ion. This effectively reduces the number of the excitation and pumping operations applied on the ion, thus leading to a smaller accumulated recoil energy.
Second, the earlier modes have higher entanglement success probabilities. As a result of these two effects, the effective phonon number prior to the $729\,$nm analysis pulse is lower in the $1.2\,$km case, which further leads to a higher detection fidelity.

Errors can also arise from the finite coherence time of the lasers and fluctuations in the optical path length,
primarily induced by the $850/854\,$nm Raman operations and $729\,$nm.
By performing Ramsey scans with these coherent operations, we observe an exponential decay time of approximately $5\,$ms (average).
Based on the time intervals between identical coherent operations in the experimental sequence, the error introduced by these operations can be roughly estimated as $\epsilon_{laser} = 0.80\%$.

Other sources of errors include ion dephasing induced by stray magnetic field, detector dark counts, mode mismatch in the BSM setup (both spatial and temporal), imperfect polarization compensation.
A list of all identified error contributions is provided in the following Table~\hyperref[table1]{S1}.

\renewcommand{\arraystretch}{1.5}
\renewcommand\tabularxcolumn[1]{m{#1}}
\begin{table}[h!]
  \begin{center}
  \small
    \begin{tabularx}{\textwidth} {
    >{\centering\arraybackslash}X
    >{\centering\arraybackslash}X
    >{\centering\arraybackslash}X}
    \hline
    \hline
      & $20\,$m case &  $1.2\,$km case\\[1pt]
    \hline
    Raman scattering & $0.98\%$   &  $0.98\%$ \\
    $866\,$nm Raman merging (induced by phase jitter) &  $0.41\%$   & $0.41\%$ \\
    Motional heating &  $0.83\%$  & $0.31\%$ \\
    Laser locking and laser fiber noise &  $0.80\%$  & $0.80\%$ \\
    Ion dephasing & $0.02\%$  & $0.17\%$ \\
    Polarisation mixing &  $0.6\%$  & $0.8\%$ \\
    Dark count & $0.1\%$  & $0.2\%$ \\
    Mode mismatch & $0.1\%$  & $0.1\%$ \\
    Miscellaneous errors & $0.76\%$  & $0.33\%$ \\
    \hline
    \hline
    \end{tabularx}
  \end{center}
  \caption{\textbf{Error budget for ion-ion entanglement}.}
  \label{table1}
\end{table}

Here we note that, in practice, several unquantifiable systematic errors also exist in our setup.
For example, temperature drifts in the laboratory can lead to fluctuations in the laser intensity used for ion operations and simultaneously cause slow polarization drifts in the optical fibers.
During the experiment, we indeed observe that the phases slowly vary over time. We categorize such uncontrolled drifts as part of the miscellaneous errors listed in the table above.
Because of the seasonal change in our city, the $1.2\,$km case exhibited a more stable thermal environment compared with the $20\,$m case.
In addition, the optical fibers in the long-distance configuration are better protected, and the phase calibrations are performed more frequently.
Therefore, it is reasonable that the long-distance case shows higher stability.
We emphasize that these systematic errors are in the scope of engineering and can be further mitigated.
In future implementations, active power stabilization within the timing sequence, improved temperature control of the experimental platform, and active polarization feedback in field-deployed fibers can be employed to suppress such drifts.
\subsection*{Section 4. BSM setup}

Photons from Alice and Bob are guided to a Bell-State Measurement (BSM) setup (see Fig.~\hyperref[FIGS3]{S3}). Mechanical shutters can block photons from one of the traps, thus allowing us to perform ion-photon experiments using the same setup. A pair of quarter-wave plate (QWP) and half-wave plate (HWP) are placed after the input fibers to compensate for stress-induced birefringence.

Following the BS, basis waveplates and polarizing beam beamsplitters (PBSs) are used to set the specific photon polarization basis in ion-photon experiments. These basis waveplates are mounted on a motorized translation stage, which can be moved out of the beam path for ion-ion experiments. The polarization imbalance of the BS are illustrated in Table~\hyperref[table2]{S2}, respectively.

\renewcommand{\arraystretch}{1.5}
\renewcommand\tabularxcolumn[1]{m{#1}}
\begin{table}[h!]
  \begin{center}
  \small
    \begin{tabularx}{\textwidth}{
    >{\centering\arraybackslash}X
    >{\centering\arraybackslash}X
    >{\centering\arraybackslash}X}
    \hline
    \hline
    50:50 BS & Reflectance & Transmission \\
    \hline
    S(V)     & $49.317\%$    & $49.977\%$     \\
    P(H)     & $49.997\%$    & $50.689\%$     \\
    \hline
    \hline
    \end{tabularx}
  \end{center}
  \caption{\textbf{The polarization imbalance of the $50$:$50$ beamsplitter}.}
\label{table2}
\end{table}

\subsection*{Section 5. Experiment control and synchronization scheme}

\begin{figure}[htbp]
    \centering
    \includegraphics[width=0.8\textwidth]{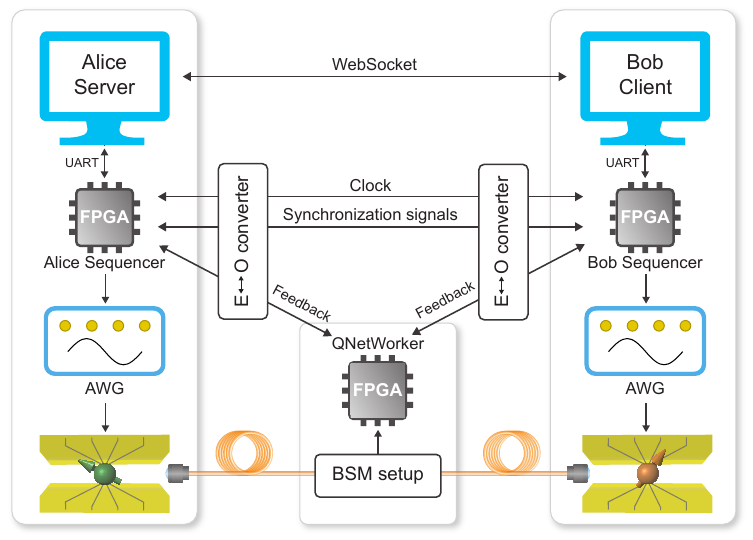}
    \caption{\textbf{Experimental control of the setup}. Sequencers in each lab execute synchronized experimental sequences. Single photons are analyzed in BSM and the outputs of SPADs are sent to QNetWorker. If the QNetWorker detects an entanglement event, it will send a signal back to Sequencers for subsequent sequences in each trap. Electrical-Optical (E-O) converters are used to transform electric and optical signals for long-distance transmission.}
    \label{FIGS4}
\end{figure}

As illustrated in Fig.~\hyperref[FIGS4]{S4}, a field programmable gate array (FPGA) `Sequencer' is developed for controlling the usual sequences and managing the overall branch of
decision to enable swift scheduling of entanglement generation and subsequent operations.
The outputs of SPADs are sent to QNetWorker for pattern matching and real-time feedback.
For ion-photon entanglement sequences, both Alice's and Bob's Sequencer can trigger the QNetWorker individually, and the QNetWorker can be connected to any one of them. However, for ion-ion entanglement sequences, the system operates with a Server/Client configuration: Alice's Sequencer serves as the Server, and Bob's Sequencer serves as the Client. The host PCs communicate with each other via the WebSocket protocol. The clock and synchronization signals are distributed over optical fibers for timing synchronization.

\end{document}